\documentclass[sigconf]{acmart}
\usepackage{graphicx}
\usepackage{float}
\usepackage{multirow}
\usepackage{booktabs}
\usepackage{subcaption}

\usepackage{amssymb} 
\usepackage{makecell}
\usepackage{array} 
\AtBeginDocument{%
  }

\setcopyright{acmlicensed}
\copyrightyear{2018}
\acmYear{2018}
\acmDOI{XXXXXXX.XXXXXXX}

\acmConference[WWW '25 Companion]{Companion Proceedings of the ACM Web Conference 2025}{April 28-May 2, 2025}{Sydney, Australia}
\acmISBN{978-1-4503-XXXX-X/18/06}

\begin{document}

%%
%% The "title" command has an optional parameter,
%% allowing the author to define a "short title" to be used in page headers.
\title{Enhancing Web Service Anomaly Detection via Fine-grained Multi-modal Association and Frequency Domain Analysis}

%%
%% The "author" command and its associated commands are used to define
%% the authors and their affiliations.
%% Of note is the shared affiliation of the first two authors, and the
%% "authornote" and "authornotemark" commands
%% used to denote shared contribution to the research.
\author{Xixuan Yang}
\orcid{0009-0008-5207-3320}
\affiliation{%
  \institution{Peking University}
  % \city{Beijing}
  \country{China}
}
\email{yangxixuan@stu.pku.edu.cn}

\author{Xin	Huang}
\affiliation{%
  \institution{Nanyang Normal University}
  % \city{Nanyang}
  % \state{Henan}
  \country{China}
}
\email{huangxin@nynu.edu.cn}

\author{Chiming	Duan}
\affiliation{%
  \institution{Peking University}
  % \city{Beijing}
  \country{China}
}
\email{duanchiming@stu.pku.edu.cn}

\author{Tong Jia}
\authornote{Corresponding Author}
\affiliation{%
  \institution{Peking University}
  % \city{Beijing}
  \country{China}
}
\email{jia.tong@pku.edu.cn}

\author{Shandong Dong}
\affiliation{%
  \institution{Alibaba group}
  % \city{Hangzhou}
  \country{China}
}
\email{dongshandong@alibaba-inc.com}

\author{Ying Li}
\affiliation{%
  \institution{Peking University}
  % \city{Beijing}
  \country{China}
}
\email{li.ying@pku.edu.cn}

\author{Gang Huang}
\affiliation{%
  \institution{Peking University}
  % \city{Beijing}
  \country{China}
}
\email{hg@pku.edu.cn}

% \author{Huifen Chan}
% \affiliation{%
%   \institution{Tsinghua University}
%   \city{Haidian Qu}
%   \state{Beijing Shi}
%   \country{China}}

%%
%% By default, the full list of authors will be used in the page
%% headers. Often, this list is too long, and will overlap
%% other information printed in the page headers. This command allows
%% the author to define a more concise list
%% of authors' names for this purpose.
\renewcommand{\shortauthors}{Xixuan Yang et al.}
\renewcommand{\shorttitle}{FFAD: Web Service Anomaly Detection}

%%
%% The abstract is a short summary of the work to be presented in the
%% article.
\begin{abstract}
Anomaly detection is crucial for ensuring the stability and reliability of web service systems. Logs and metrics contain multiple information that can reflect the system's operational state and potential anomalies. Thus, existing anomaly detection methods use logs and metrics to detect web service systems' anomalies through data fusion approaches. They associate logs and metrics using coarse-grained time window alignment and capture the normal patterns of system operation through reconstruction. However, these methods have two issues that limit their performance in anomaly detection. First, due to asynchrony between logs and metrics, coarse-grained time window alignment cannot achieve a precise association between the two modalities. Second, reconstruction-based methods suffer from severe overgeneralization problems, resulting in anomalies being accurately reconstructed. In this paper, we propose a novel anomaly detection method named FFAD to address these two issues. On the one hand, FFAD employs graph-based alignment to mine and extract associations between the modalities from the constructed log-metric relation graph, achieving precise associations between logs and metrics. On the other hand, we improve the model's fit to normal data distributions through Fourier Frequency Focus, thereby enhancing the effectiveness of anomaly detection. We validated the effectiveness of our model on two real-world industrial datasets and one open-source dataset. The results show that our method achieves an average anomaly detection F1-score of 93.6\%, representing an 8.8\% improvement over previous state-of-the-art methods.
\end{abstract}

%%
%% The code below is generated by the tool at http://dl.acm.org/ccs.cfm.
%% Please copy and paste the code instead of the example below.
%%


%%
%% Keywords. The author(s) should pick words that accurately describe
%% the work being presented. Separate the keywords with commas.
%% A "teaser" image appears between the author and affiliation
%% information and the body of the document, and typically spans the
%% page.

% \received{20 February 2007}
% \received[revised]{12 March 2009}
% \received[accepted]{5 June 2009}

%%
%% This command processes the author and affiliation and title
%% information and builds the first part of the formatted document.
\maketitle

\section{INTRODUCTION}
Web services are software systems that support interoperable machine-to-machine interactions over a network using standardized protocols such as HTTP, SOAP, and REST. They are critical in modern distributed systems, enabling different applications to communicate and exchange data seamlessly. However, the complex interactions and dynamic environments in which web services operate pose significant challenges to maintaining their stability and reliability. Therefore, timely anomaly detection is crucial for preventing service interruptions and ensuring the reliability of web service systems, which has become a widely studied topic in recent years.

Existing anomaly detection methods are usually based on two modalities of data: logs and metrics. Logs record important events during system operation, while metrics record the performance indicators of web services. Due to the complex service interactions and diverse software and hardware environments involved in web service systems, single-modal data often cannot capture all types of anomalies \cite{zhao2021identifying, lee2023heterogeneous,liu2024uac}. Therefore, existing methods usually adopt a multi-modal fusion approach, combining data from both logs and metrics for anomaly detection. Specifically, existing methods usually align logs and metrics based on coarse-grained time window alignment, associating them within the same time window and detecting anomalies by reconstructing the normal patterns of the system operation.

\begin{figure}[ht]
    \centering
    \includegraphics[width=1.0\linewidth]{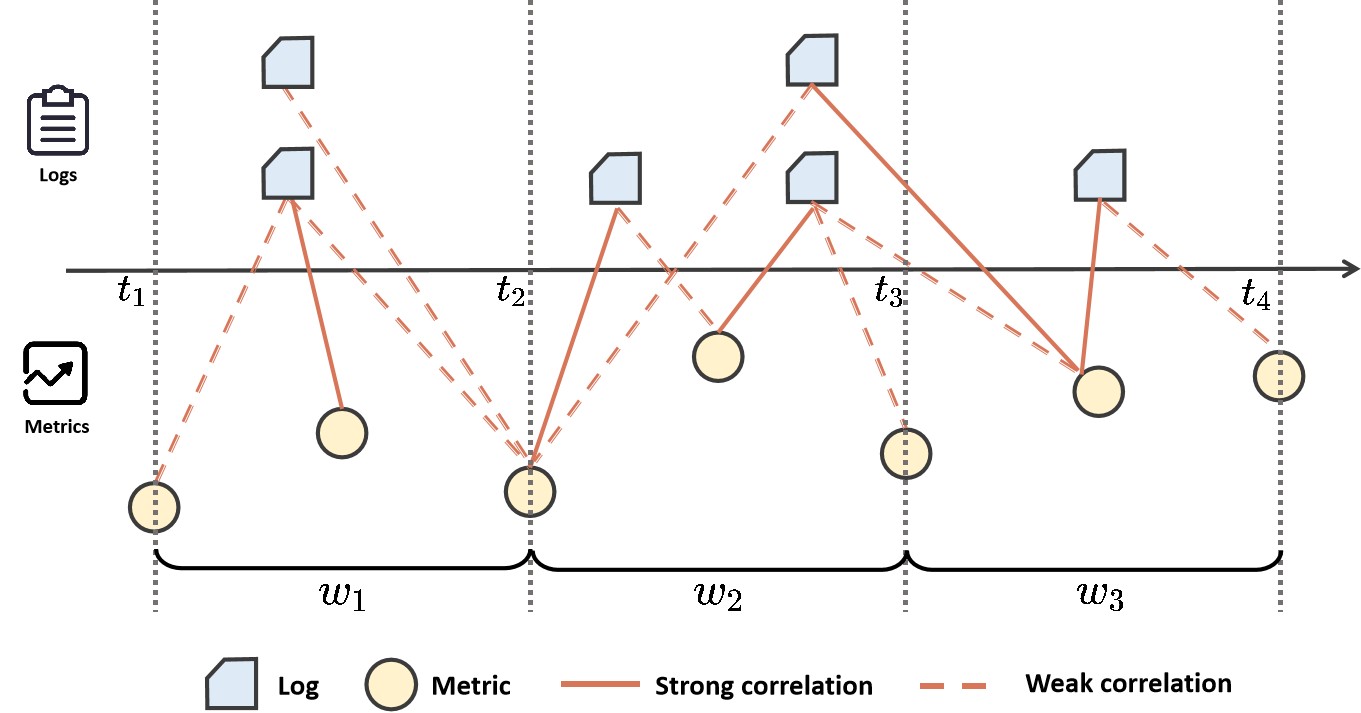} 
    \caption{Precise Association Between Logs and Metrics}
    \label{fig:intro}
    \vspace{-0.5cm}
\end{figure}

However, existing anomaly detection methods have two problems. Coarse-grained time window alignment methods cannot achieve precise associations between log and metric data. Due to processing delays, network delays, and asynchronous communication among components of web service systems, there is generally asynchrony between logs and metrics. As shown in Figure \ref{fig:intro}, even within the same time window, different log entries may correspond to metric sampling points at different timestamps. Simply aligning by time window can only associate batches of log entries and metric sampling points and cannot accurately model their precise associations. In practical industrial scenarios, there are numerous log entries and metric sampling points within the same time window, and the associations among them are very complicated, making it difficult for existing methods to handle them effectively. This poses a significant obstacle to accurately modeling system states and identifying anomalies.

Another issue is that reconstruction-based methods have serious overgeneralization problems, where anomalous inputs can also be well reconstructed, leading to limited accuracy in anomaly detection. This occurs because the encoder extracts unique features of an anomaly or the decoder has excessive decoding capabilities for abnormal encoding vectors \cite{park2020learning,song2023memto}.

In this paper, we propose \textbf{FFAD} (\textbf{F}ine-grained multi-modal Association and \textbf{F}requency Domain Analysis for \textbf{A}nomaly \textbf{D}etection), a novel multi-modal anomaly detection method, to address the above two issues. FFAD uses a graph-based approach to align logs and metrics, achieving a fine-grained and precise association between these modalities. We treat each log entry and metric sampling point within a time window as a node in the graph. We mine all possible associations between them by constructing a fully connected graph using data within a sliding window. Subsequently, we assign weights to each edge in the log-metric relation graph to quantitatively distinguish the importance levels of different association relationships. To reduce the training cost at this stage, inspired by FourierGNN \cite{yi2024fouriergnn}, we adopt the Fourier Graph Operator (FGO) to replace traditional graph computation units (such as convolutions), performing matrix multiplication in the Fourier space of the graph to capture the deep associations between logs and metrics effectively. Moreover, the amount of data in different modalities can affect the extraction of associations between them, causing the model to often focus excessively on modalities with larger amounts of data. This disparity is especially evident in our relational graph structure, where some modalities occupy too few nodes, limiting their influence in alignment. We inject specific types of noise into the data-rich modalities based on their modal features to balance the learning between logs and metrics and prevent the model from over-relying on modalities with a larger amount of data.

In addition, we propose the Fourier Frequency Focus strategy to address the overgeneralization problem of reconstruction-based methods. Specifically, we calculate the signal strength and amplitude of each frequency component, dynamically reducing the complex values of frequency components with high signal strength (which may contain anomalous information) during matrix operations. This allows the model to focus more on learning normal patterns. During the reconstruction process, this mechanism amplifies the reconstruction errors of anomalous points, thereby improving the ability to detect anomalies.

We conducted extensive experiments on three datasets, including two real industrial datasets. The results show that our method achieves an average F1-score of 93.6\% for anomaly detection, an improvement of 8.8\% over previous state-of-the-art methods. FFAD has been successfully deployed in production since September 2023, serving the Taobao platform in Alibaba.

The main contributions of this work are summarized as follows:

\begin{itemize}
    \item We propose FFAD, a new multi-modal anomaly detection method using logs and metrics, which addresses two major problems of existing methods and improves the accuracy of anomaly detection tasks.
    \item FFAD uses a graph-based alignment method, assigns weights based on the importance of association relationships from the constructed log-metric relational graph using FGO, and balances each modality through dual noise injection, achieving precise association of logs and metrics.
    \item FFAD addresses the overgeneralization problem of recon-struction-based methods through the Fourier Frequency Focus strategy, improving the model's fit to the normal data distribution and thereby enhancing the effectiveness of anomaly detection.
    \item We conducted comprehensive experiments on three datasets, and FFAD achieved state-of-the-art results.
\end{itemize}

\section{RELATED WORK}

In recent years, anomaly detection has received significant attention and has been extensively studied to ensure the reliability of large-scale systems \cite{chen2024lara,jie2024disentangled,tan2024air,zhao2024weakly}. Anomaly detection methods are typically based on logs, metrics, or a combination of both. Metric-based methods can be categorized into three types. Density estimation-based methods assume that normal data follow a specific probability distribution and identify anomalies by evaluating the probability densities of data points \cite{breunig2000lof,tang2002enhancing,zong2018deep}. Clustering-based methods treat outliers as anomalies and identify them based on the distances between data points and cluster centres \cite{tax2004support,shen2020timeseries}. Reconstruction-based methods posit that anomalous data are difficult to reconstruct, and anomalies are determined through reconstruction errors \cite{park2018multimodal,su2019robust,zhan2022stgat}. However, these methods have limitations, such as assuming specific data distributions and heavily relying on hyperparameter configurations, which make them challenging to deploy in real production environments.

Log-based methods utilize semi-structured text collected at the application or system level for anomaly detection. Traditional approaches detect anomalies by identifying keywords like "error" or "fail" or counting the number of logs, but these offer limited accuracy. Advanced methods involving log parsing, feature extraction, and anomaly detection have been proposed to address these limitations \cite{xu2010system,lin2016log,guan2023grasped,fu2023mlog}. These methods aim to uncover log patterns of normal executions and detect anomalies when current executions deviate from these patterns. However, log-based methods that rely on a single data source often struggle to meet the precision demands of practical applications \cite{zhao2021identifying, lee2023heterogeneous,liu2024uac}.

Multi-modal methods attempt to comprehensively analyze multiple data sources, such as logs and metrics, to reveal anomalies more thoroughly. Zhao et al. \cite{zhao2021identifying} use LSTM networks to encode log and metric sequences, fusing features through concatenation or cross-attention. Liu et al. \cite{liu2024uac} combine adversarial and contrastive learning to capture complex patterns and enhance discrimination between normal and abnormal samples. Lee et al. \cite{lee2023heterogeneous} presented a semi-supervised method that employs FastText for log embedding, Transformers for log feature extraction, CNNs for metric feature extraction, and integrates features via concatenation. These methods are all reconstruction-based. However, existing anomaly detection methods have two problems. First, coarse-grained time window alignment methods cannot achieve precise associations between log and metric data. Second, reconstruction-based methods have serious overgeneralization problems, where anomalous inputs can also be well reconstructed, leading to limited accuracy in anomaly detection.

\section{METHODOLOGY}

As previously discussed, existing anomaly detection methods in web service systems have two major problems. First, coarse-grained time window alignment methods cannot achieve precise associations between log and metric data. Second, reconstruction-based methods have serious overgeneralization problems, where anomalous inputs can also be well reconstructed, leading to limited accuracy in anomaly detection. We propose FFAD, a novel multi-modal anomaly detection method, to address the above two issues. FFAD uses a graph-based approach to align logs and metrics, achieving a fine-grained and precise association between these modalities. In addition, we propose the Fourier Frequency Focus strategy to address the overgeneralization problem of reconstruction-based methods.

FFAD comprises five main modules: data preprocessing, temporal feature retention, dual noise injection module, graph processing, and output projection. Benefiting from the unique advantages of these modules, our method can effectively fuse multi-modal data, balance data volume differences between modalities, capture anomaly patterns, and improve the accuracy of anomaly detection.

\begin{figure*}[ht]
  \centering
  \includegraphics[width=\textwidth]{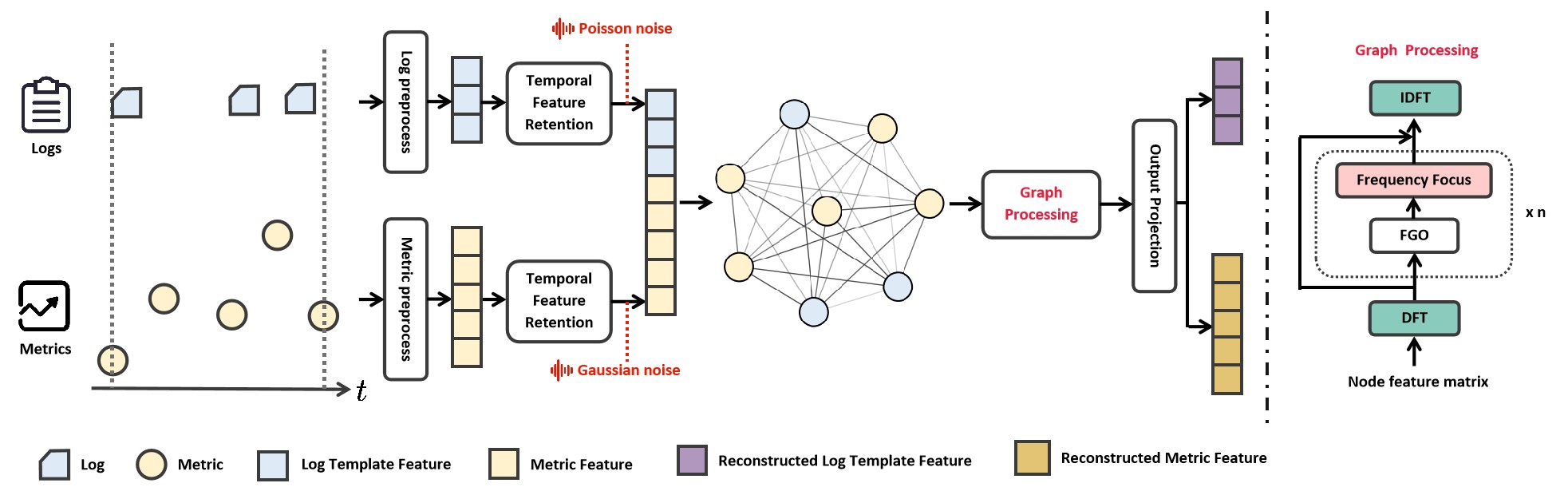}
  \caption{The overview of FFAD.}
  \label{overview}
  \vspace{-0.5cm}
\end{figure*}

Firstly, we perform data preprocessing to unify the structures of metric data and log data by constructing metric sequences and log template sequences, respectively. These sequences are then input into the temporal feature retention module, which performs convolution operations along the time dimension on the metric sequences and log template sequences separately to capture the temporal dependencies of the data. Considering the imbalance in data volume between modalities, we propose the Dual Noise Injection module, which injects specific types of noise into the modality with a larger data volume to balance the model's learning attention across different modalities. Subsequently, we construct a fully connected fusion graph, representing all data points from both modalities as nodes in the graph to model cross-modal interaction relationships. However, processing such a large-scale graph structure has high computational complexity. To address this, we perform graph processing in the Fourier space of the graph. Furthermore, we propose the \textit{Fourier Frequency Focusing} strategy, which focuses on low-energy frequency components in the Fourier space to suppress the influence of anomalous signals on the model and enhance the model's ability to capture normal patterns. Finally, the features are sent to the output projection module to obtain the reconstruction of the input data. We can effectively detect potential anomalies by comparing the differences between the reconstructed data and the original data. Figure~\ref{overview} illustrates the overall workflow of FFAD.

\subsection{Data preprocessing}
Given a series of metrics and log data from a web service system, metric data adopts a typical time-series format, which is easy to process. However, log data presents difficulties for analysis due to its semi-structured or unstructured nature and the large number of variables it contains. To facilitate the construction of the subsequent fully connected graph, we unify the structures of the log data and metric data. Specifically, we define metric sequences ${ m(t) }$ and log template sequences ${ l(t) }$ sampled at fixed time intervals $\Delta t$.

For the metric data, at each time point $t$, we collect metric values to generate a metric vector $m(t) = [m_1(t), m_2(t), \ldots, m_n(t)]$, where $m_j(t)$ represents the value of metric $M_j$ at time $t$, and $n$ is the total number of metrics.

For the log data, we employ the widely used parser Drain~\cite{he2017drain} for log parsing, as it has been proven effective in previous evaluations~\cite{liu2024uac}. Specifically, we first use Drain to extract $n'$ log templates ${ E_1, E_2, \ldots, E_{n'} }$. When a log message arrives, we match it with the corresponding template, converting it into a log template sequence arranged in chronological order. Studies have shown that log templates exhibit significant differences between normal and abnormal system states: frequently occurring templates have lower discriminative power, while rare templates containing error handling are more valuable for detection~\cite{liu2024uac, lin2016log}. Therefore, we focus on the occurrence of each template, treating multiple identical log templates within the same time block as a single occurrence to highlight those rare log events that may be related to anomalies. For each time block $[t, t + \Delta t)$, we record whether each log template has appeared. In this way, we generate a log vector of length $n'$ for each time block, $l(t) = [l_1(t), l_2(t), \ldots, l_{n'}(t)]$, where $l_k(t) \in { 0, 1 }$ indicates whether template $E_k$ appears in time block $t$.

Based on the above definitions, we construct multi-modal time-series data $X = { x_t }{t=1}^T$, where $x_t = [ m(t), l(t) ]$ combines the metric and log vectors at time $t$, and $T$ denotes the total number of time blocks. The goal of anomaly detection is to determine whether $X^{(i)}$ contains anomalies, where $X^{(i)} = { x_t }{t=i-w+1}^i$ represents the multi-modal data in the $i$th sliding window, and $w$ is the size of the sliding window.

\subsection{Temporal Feature Retention}
Fully connected graphs can capture global associations between nodes, but they ignore the temporal order among nodes during construction. In microservice systems, both metric and log data exhibit strong temporal correlations; if temporal features are not considered, it may not be possible to fully uncover anomalies in the data. We design a Temporal Feature Retention (TFR) module, which performs convolution along the time dimension on the metric and log template sequences separately. For the multi-modal data \( X^{(i)} = \{ x_t \}_{t=i-w+1}^i \) in the \( i \)th sliding window, we input the metric sequence \( \{ m(t) \}_{t=i-w+1}^i \) into the temporal feature retention module \( \text{TFR}_m \), obtaining the metric features \( m'^{(i)} \in \mathbb{R}^{w \times d_m} \) that incorporate temporal information:

\begin{equation}
m'^{(i)} = \text{ReLU}\left( \text{Conv1D}_m\left( \text{Pad}\left( \left\{ m(t) \right\}_{t=i-w+1}^i \right) \right) \right)
\end{equation}
Similarly, we input the log template sequence \( \{ l(t) \}_{t=i-w+1}^i \) into the temporal feature retention module \( \text{TFR}_l \), obtaining the log template features \( l'^{(i)} \in \mathbb{R}^{w \times d_l} \) that incorporate temporal information:
\begin{equation}
l'^{(i)} = \text{ReLU}\left( \text{Conv1D}_l\left( \text{Pad}\left( \left\{ l(t) \right\}_{t=i-w+1}^i \right) \right) \right)
\end{equation}

\subsection{Dual Noise Injection}
In web service systems, the data volumes of different modalities often vary significantly. This imbalance is particularly evident in fully connected graph structures: modalities with smaller data volumes constitute a disproportionately low number of nodes in the graph, thereby limiting their influence in modality fusion. To address this issue, we propose a Dual Noise Injection(DNI) Mechanism, which balances the model's learning attention to different modalities by injecting specific types of noise into the data. Specifically, for continuous metric data, we introduce Gaussian noise to simulate random fluctuations; for discrete log data, we employ Poisson noise, suitable for modeling the randomness of discrete events. Based on the relative magnitudes of the metric data and log data, we inject different types of noise accordingly. The perturbed metric features $\tilde{m}'^{(i)}$ and log template features $\tilde{l}'^{(i)}$ are defined as: 

\begin{equation}
\tilde{m}'^{(i)} =
\begin{cases}
m'^{(i)} + \alpha_m \epsilon_m^{(i)}, & \text{if } D_m \gg D_l,\ \text{where } \epsilon_m^{(i)} \sim \mathcal{N}(0, \sigma^2 I_{d_m}) \\
m'^{(i)}, & \text{otherwise}
\end{cases}
\end{equation}
\begin{equation}
\tilde{l}'^{(i)} =
\begin{cases}
l'^{(i)} + \alpha_l \epsilon_l^{(i)}, & \text{if } D_l \gg D_m,\ \text{where } \epsilon_l^{(i)} \sim \text{Pois}(\lambda \cdot \mathbf{1}{d_l}) \\
l'^{(i)}, & \text{otherwise}
\end{cases}
\end{equation}

, where $\epsilon_m^{(i)}$ is a Gaussian noise vector with standard deviation $\sigma$, $\epsilon_l^{(i)}$ is a Poisson noise vector with parameter $\lambda$, $\mathbf{1}{d_l}$ is an all-ones vector of length $d_l$, $D_m$ and $D_l$ are the data volumes of the metrics and logs, respectively, and $\alpha_m$ and $\alpha_l$ are noise scaling coefficients. The core idea of DNBM is to increase the learning difficulty of the modality with a larger data volume, thereby prompting the model to pay more attention to the modality with less data. By introducing noise, we increase the reconstruction error of the modality with more data, forcing the model during training to better utilize information from the modality with less data to reduce the overall reconstruction error.

\subsection{Fully-Connected Fusion Graph }
To achieve precise association of logs and metrics, we propose constructing a fully connected fusion graph \( \mathcal{G}_i \). This graph uniformly represents all data points from both modalities as nodes, thereby comprehensively modeling cross-modal interaction relationships. Specifically, for the \( i \)th sliding window, we flatten the processed metric features and log template features separately. For the metric features, we obtain \( X_m(i) = \text{vec}\left( (\tilde{m}'^{(i)})^\top \right) \in \mathbb{R}^{ N_m \times 1 } \), where \( N_m = w \times d_m \). For the log features, we obtain \( X_l(i) = \text{vec}\left( (\tilde{l}'^{(i)})^\top \right) \in \mathbb{R}^{ N_l \times 1 } \), where \( N_l = w \times d_l \). We then concatenate these two flattened vectors to form the node feature vector \( X_i^\mathcal{G} = [ X_m(i); X_l(i) ] \in \mathbb{R}^{ N \times 1 } \), where \( N = N_m + N_l \). Each node corresponds to a specific feature at a specific time step within the sliding window. Next, we construct a fully connected fusion graph \( \mathcal{G}_i = ( V_i, E_i ) \), where the node set \( V_i \) contains \( N \) nodes corresponding to each element in \( X_i^\mathcal{G} \). The edge set \( E_i \) connects every pair of nodes, represented by the adjacency matrix \( A_i^\mathcal{G} \in \{ 1 \}^{ N \times N } \), where all elements are \( 1 \). Finally, we map the node feature vector \( X_i^\mathcal{G} \) to node embeddings through an embedding function \( \phi \), that is, \( H_i^\mathcal{G} = \phi( X_i^\mathcal{G} ) \in \mathbb{R}^{ N \times d' } \), where \( d' \) is the dimension of the embedding space.

\subsection{Graph Processing}
As the number of sequences grows and the window length expands, the number of nodes in a fully connected graph increases, potentially making the computational cost of classical graph networks (usually with quadratic complexity) prohibitively high. Inspired by FourierGNN~\cite{yi2024fouriergnn}, we replace the standard graph operations (e.g., convolution) with the Fourier Graph Operator (FGO), performing matrix multiplication in the Fourier space of the graph.

Moreover, we observe that many reconstruction-based methods suffer from overgeneralization, resulting in overly accurate reconstruction of anomalous inputs~\cite{park2020learning,song2023memto}. Studies have shown that, in the frequency domain, anomalies often manifest as strong signals with high-energy components, leading to greater variability in their spectra~\cite{ma2021jump,chen2024learning}. To enhance the model's ability to learn normal features, we propose a Fourier Frequency Focus (FFF) strategy, which focuses on low-energy frequency components in the Fourier space, thereby reducing the impact of abnormal signals and improving the model's ability to capture normal patterns.

Specifically, for the \( i \)th sliding window, we first perform a Discrete Fourier Transform (DFT) on the node feature matrix of the fully connected fusion graph, \( X_i^\mathcal{G} \in \mathbb{R}^{ N \times d' } \), obtaining frequency domain features \( \hat{X}_i^\mathcal{G} = \mathcal{F}( X_i^\mathcal{G} ) \). We then calculate the signal energy \( E \) and amplitude variance \( V \) of each frequency component to capture possible anomaly patterns. The signal energy is defined as the sum of squared magnitudes of the frequency domain features along the feature dimension:
\begin{equation}
E_k = \sum_{ j = 1 }^{ d' } \left| \hat{X}_i^\mathcal{G}( k, j ) \right|^2,
\end{equation}
and the amplitude variance is calculated as the variance of the magnitudes of the frequency domain features along the feature dimension:
\begin{equation}
V_k = \text{Var}\left( \left\{ \left| \hat{X}_i^\mathcal{G}( k, j ) \right| \right\}_{ j = 1 }^{ d' } \right).
\end{equation}
Here, \( \hat{X}_i^\mathcal{G}( k, j ) \) denotes the value of the \( k \)th frequency component in the \( j \)th feature dimension. Next, we set energy threshold \( E_{\text{th}} \) and variance threshold \( V_{\text{th}} \). For frequency components with energy above the threshold and large amplitude variance (i.e., components that may contain anomalous information), we introduce a learnable scaling factor \( \alpha_{\text{anomaly}} < 1 \) to scale them; for other frequency components, we keep them unchanged, using \( \alpha_{\text{normal}} = 1 \). The scaling factor is calculated as follows:
\begin{equation}
\alpha_k = 
\begin{cases} 
\alpha_{\text{anomaly}}, & \text{if } E_k > E_{\text{th}} \text{ and } V_k > V_{\text{th}}, \\ 
\alpha_{\text{normal}}, & \text{otherwise}.
\end{cases}
\end{equation}
In the frequency domain, we perform matrix multiplication \( \hat{X}_i^\mathcal{G} S_{ A_i^\mathcal{G}, W } \) using the Fourier Graph Operator \( S_{ A_i^\mathcal{G}, W } \). The operator \( S_{ A_i^\mathcal{G}, W } \) is obtained via the Discrete Fourier Transform (DFT), expressed as \( S_{ A_i^\mathcal{G}, W } = \mathcal{F}\left( \kappa( A_i^\mathcal{G}, W ) \right) \), where \( \kappa( A_i^\mathcal{G}, W ) \) is a kernel function defined based on the adjacency matrix \( A_i^\mathcal{G} \) and the weight matrix \( W \)~\cite{yi2024fouriergnn}. Matrix multiplication with the FGO in the Fourier space is equivalent to convolution operations in the time domain. We perform recursive matrix multiplication and cumulative summation; after each layer's output, we apply the scaling factor \( \alpha_k \) to continuously suppress the influence of anomalous frequency components. Specifically, the output after stacking \( q \) layers is:
\begin{equation}
H_i^{(q)} = \sigma\left( \left( H_i^{(q-1)} S_{ A_i^\mathcal{G}, W_q } + b_q \right) \odot \alpha_k \mathbf{1}_{ d' }^\top \right),\quad H_i^{(0)} = \hat{X}_i^\mathcal{G},
\end{equation}
where \( \sigma \) is a nonlinear activation function, \( b_q \) is a bias term, and \( \odot \) denotes element-wise multiplication. Finally, we convert the frequency domain features \( H_i^{(q)} \) back to time domain features \( Z_i \) through the inverse Discrete Fourier Transform (IDFT).

\subsection{Output Projection}
To reconstruct the data, we map the fused features \( Z_i \in \mathbb{R}^{ N \times d' } \) back to the input space. Specifically, we reshape \( Z_i \) into \( \bar{Z}_i \in \mathbb{R}^{ w \times s } \), where \( s = ( d_m + d_l ) d' \). Next, we apply a linear transformation to map \( \bar{Z}_i \) to \( \hat{X}_i \in \mathbb{R}^{ w \times ( d_m + d_l ) } \). Finally, we split \( \hat{X}_i \) into the reconstructed metric features \( \hat{m}_i \in \mathbb{R}^{ w \times d_m } \) and the reconstructed log template features \( \hat{l}_i \in \mathbb{R}^{ w \times d_l } \).

\section{EVALUATION}
We evaluate FFAD by answering the following research questions (RQs):

\begin{itemize}
    \item RQ1: How effective is FFAD in anomaly detection task?
    \item RQ2: How do different modules contribute to FFAD?
    \item RQ3: Does data from each modality contribute to FFAD?
    \item RQ4: How is the effectiveness of the FFF mechanism visualized in FFAD?
\end{itemize}

\subsection{Experiment Setup}
\subsubsection{Datasets}
To evaluate the performance of FFAD, we conducted extensive experiments using two open-source datasets and a real-world industrial dataset from Alibaba, as summarized in Table \ref{evaluation_datasets}. 

Dataset A is a publicly released in-lab dataset from~\cite{lee2023heterogeneous}, specifically designed to assess the effectiveness of multi-modal anomaly detection. This dataset includes 11 metrics sampled every 10 seconds. Its log data contains 117 log templates.

Dataset B is a complex multi-modal dataset simulated by an intelligent operations company. It serves as a benchmark for evaluating multi-modal anomaly detection under complex scenarios. The dataset consists of metrics and log data from four different services, including 85 metrics sampled every 30 seconds. Its log data comprises 20 log templates. This dataset incorporates various artificially injected faults, such as CPU overload, memory overload, logging failures, and network failures.

Dataset C was collected from Alibaba's real-world industrial software platform and is based on the open-source service system Train-Ticket\cite{train-ticket}. Train-Ticket is a benchmark service system that provides railway ticket services, allowing users to perform operations such as ticket booking, payment, and rescheduling. We used Locust\cite{locust} to simulate user requests, covering functionalities such as route selection and ticket booking. The concurrent user request count was set to 10, with a startup speed of 1 request per second. We collected metric data using Alibaba Cloud's Application Real-Time Monitoring Service (ARMS) and log data using Alibaba Cloud's Log Service (SLS). The dataset includes 41 metrics, such as CPU usage and memory usage, as well as log data containing 1,895 unique templates. We injected 16 types of faults, with each fault injection lasting 5 to 10 minutes.

\begin{table}[h]
\centering
\caption{The Overview of the Evaluation Datasets.}
\label{evaluation_datasets}
\begin{tabular}{|c|c|c|c|}
\hline
\textbf{Dataset} & \textbf{Log Message} & \textbf{Metric Length} & \textbf{Anomaly Ratio} \\
\hline
Dataset A & 11,141,400 & 55,707 & 26.00\% \\
\hline
Dataset B & 5,308,847 & 42,813 & 5.63\% \\
\hline
Dataset C & 344,510 & 7,242 & 23.44\% \\
\hline
\end{tabular}
\end{table}

\subsubsection{Baselines}
To validate the effectiveness of our method, we conducted comprehensive comparisons with various baseline methods. These comparisons involved different data sources, such as logs, metrics, and multi-modal data, as well as different training strategies, including supervised, semi-supervised, and unsupervised learning paradigms. Next, we will introduce these methods.

Deeplog~\cite{du2017deeplog} is an unsupervised log anomaly detection method that utilizes LSTM networks to encode log sequences and predict the next log template. An anomaly is detected if the next template is not among the top-$k$ predicted templates. LogRobust~\cite{zhang2019robust} is a supervised log anomaly detection method that encodes log sequence features and performs classification to achieve robust anomaly detection performance. OmniAnomaly~\cite{su2019robust} is an unsupervised metric-based anomaly detection method that integrates Variational Autoencoders (VAE) into time series analysis to compress features and learn normal pattern distributions. MTAD-GAT~\cite{zhao2020multivariate} is an unsupervised metric-based anomaly detection method that employs Graph Attention Networks (GAT) to capture spatial correlations in multivariate time series. SCWarn~\cite{zhao2021identifying} is an unsupervised multi-modal anomaly detection method that uses LSTM networks to encode log and metric sequences, fusing features through concatenation or cross-attention operations. Hades~\cite{lee2023heterogeneous} is a semi-supervised multi-modal anomaly detection method that employs FastText~\cite{joulin2016fasttext} for log embedding, Transformers for log feature extraction, CNNs for metric feature extraction, and integrates features via concatenation. UAC-AD~\cite{liu2024uac} is an unsupervised multi-modal anomaly detection method combining adversarial and contrastive learning to capture complex patterns between logs and metrics, enhancing discrimination by increasing separation between normal and abnormal samples.

\subsubsection{Evaluation Metrics}
To comprehensively assess the effectiveness of our anomaly detection method, we use Precision, Recall, and F1-score as evaluation metrics. They are defined as follows: $\text{Precision (Pre)} = \frac{TP}{TP + FP}$, $\text{Recall (Rec)} = \frac{TP}{TP + FN}$, and $\text{F1-score} = \frac{2 \times \text{Pre} \times \text{Rec}}{\text{Pre} + \text{Rec}}$, where TP, FP, and FN denote true positives, false positives, and false negatives, respectively.

\subsubsection{Implementation Details}
To ensure the reproducibility of our research, we provide comprehensive implementation details. Following the methodology in \cite{liu2024uac}, we use Drain as the log parser and group log and metric sequences using a sliding window of size 50. We split the dataset into training, testing, and validation sets in a ratio of 7:2:1. We optimize FFAD using the Adam optimizer with a learning rate of $5 \times 10^{-4}$ and an initial batch size of 256. Based on the relative scales of metric data and log data in different datasets, we adjusted the noise ratio coefficients $\alpha_m$ and $\alpha_l$. Notably, the amount of log data refers to the number of log templates after parsing, rather than the original log data size. Specifically, we varied $\alpha_m$ and $\alpha_l$ over the range {0,1,2,3,4,5,6,7,8,9}. For Dataset A, we set $\alpha_m = 0$ and $\alpha_l = 1$; for Dataset B, $\alpha_m = 1$ and $\alpha_l = 0$; for Dataset C, $\alpha_m = 0$ and $\alpha_l = 7$. We adjusted the variance of the injected noise, selecting from candidate values of {0.007, 0.009, 0.01, 0.03, 0.05, 0.07, 0.1}. We set the noise variance to 0.007 for Dataset A, and 0.003 for Datasets B and C. We set the node embedding dimension $d'$ to 128. To focus on important frequency components, we set the energy threshold $E_{\text{th}}$ and variance threshold $V_{\text{th}}$ at the 95th percentile of their respective distributions, thereby focusing only on the top 5\% frequency components with the highest energy and variance. We set the number of stacked computational layers $q$ in the Fourier space to 3. All experiments were conducted on a Linux server equipped with an Intel Xeon(R) Silver 4214R 2.4GHz processor and an RTX3090 GPU with 24GB of memory. We used the publicly available implementations and parameter settings of baseline models for comparison, including \cite{du2017deeplog,zhang2019robust,zhang2019robust,su2019robust,zhao2020multivariate,zhao2021identifying,lee2023heterogeneous,liu2024uac}.

\subsection{RQ1: How effective is FFAD in anomaly detection task?}

\begin{table*}[ht]
    \centering
    \caption{Overall Performance Comparison of FFAD and Several State-of-the-Art Methods}
    \begin{tabular}{lll ccc ccc ccc c}
        \toprule
        \multirow{2}{*}{Methods} 
        & \multirow{2}{*}{Source} 
        & \multirow{2}{*}{Training Strategy} 
        & \multicolumn{3}{c}{Dataset A} 
        & \multicolumn{3}{c}{Dataset B} 
        & \multicolumn{3}{c}{Dataset C} 
        & Avg. \\ % 将 "Avg." 放在第一行
        \cmidrule(lr){4-6} 
        \cmidrule(lr){7-9} 
        \cmidrule(lr){10-12} 
        \cmidrule(lr){13-13} 
        & & 
        & F1  & Rec  & Pre  
        & F1  & Rec  & Pre  
        & F1  & Rec  & Pre  
        & F1 \\ 
        \midrule
        DeepLog                  & log                     & unsupervised                       & 0.250 & 0.373 & 0.189 & 0.560 & 0.430 & 0.805 & 0.223 & 0.139 & 0.559 & 0.344 \\
        LogRobust                & log                     & supervised                         & 0.835 & 0.840 & 0.830 & 0.775 & 0.835 & 0.725 & 0.397 & 0.289 & 0.633 & 0.669 \\
        OmniAnomaly              & metric                  & unsupervised                       & 0.690 & 0.790 & 0.612 & 0.280 & 0.445 & 0.205 & 0.254 & 0.225 & 0.291 & 0.408 \\
        MTAD-GAT                 & metric                  & unsupervised                       & 0.730 & 0.770 & 0.695 & 0.390 & 0.475 & 0.330 & 0.370 & \underline{0.869} & 0.235 & 0.497 \\
        SCWarn                   & multi-modal             & unsupervised                       & 0.800 & 0.880 & 0.735 & 0.400 & 0.545 & 0.317 & 0.471 & 0.382 & 0.615 & 0.557 \\
        Hades                    & multi-modal             & semi-supervised                    & \underline{0.918} & \underline{0.944} & 0.895 & 0.810 & 0.830 & 0.790 & \underline{0.773} & 0.735 & \underline{0.815} & 0.834 \\
        UAC-AD                   & multi-modal             & unsupervised                       & 0.878 & 0.858 & \underline{0.900} & \underline{0.956} & \underline{0.958} & \textbf{0.956} & 0.711 & 0.676 & 0.750 & \underline{0.848} \\
        FFAD (Ours)                   & multi-modal             & unsupervised                       & \textbf{0.934} & \textbf{0.965} & \textbf{0.904} & \textbf{0.961} & \textbf{1.0} & \underline{0.925} & \textbf{0.914} & \textbf{1.0} & \textbf{0.841} & \textbf{0.936} \\ 
        \bottomrule
    \end{tabular}
    \label{tab:performance_comparison}
    \vspace{-0.2cm}
\end{table*}

\begin{table*}[ht]
    \centering
    \caption{Ablation Study on Each Component of FFAD Over Three Datasets}
    \begin{tabular}{c ccc ccc ccc ccc c}
        \toprule
        \multirow{2}{*}{Method} & 
        \multicolumn{3}{c}{Strategies} & 
        \multicolumn{3}{c}{Dataset A} & 
        \multicolumn{3}{c}{Dataset B} & 
        \multicolumn{3}{c}{Dataset C} & 
        Avg. \\   % 第一行表头
        \cmidrule(lr){2-4} 
        \cmidrule(lr){5-7} 
        \cmidrule(lr){8-10}
        \cmidrule(lr){11-13} 
        \cmidrule(lr){14-14} % 为最后一列添加横线
        & TFR & DNI & FFF & F1  & Rec  & Pre  & F1  & Rec  & Pre  & F1  & Rec  & Pre  & F1 \\ % 第二行表头
        \midrule
        Method 1 &  & \checkmark & \checkmark & 0.910 & 0.966 & 0.860 & 0.917 & 1.0   & 0.847 & 0.825 & 0.835 & 0.815 & 0.884 \\
        Method 2 & \checkmark &  & \checkmark & 0.906 & 0.964 & 0.856 & 0.864 & 0.968 & 0.782 & 0.870 & 1.0   & 0.770 & 0.880 \\
        Method 3 & \checkmark & \checkmark &  & 0.908 & 0.929 & 0.887 & 0.923 & 0.988 & 0.867 & 0.873 & 1.0   & 0.774 & 0.901 \\
        Method 4 (Ours) & \checkmark & \checkmark & \checkmark & \textbf{0.934} & 0.965 & 0.904 & \textbf{0.961} & 1.0 & 0.925 & \textbf{0.914} & 1.0 & 0.841 & \textbf{0.936} \\
        \bottomrule
    \end{tabular}
    \label{tab:ablation_results}
    \vspace{-0.2cm}
\end{table*}

In this section, we evaluate the effectiveness of FFAD in anomaly detection tasks. As shown in Table \ref{tab:performance_comparison}, we compare our method with various baseline models, including unimodal methods, multi-modal methods, semi-supervised methods, and unsupervised methods.

Overall, FFAD achieves new state-of-the-art performance on all datasets. Our method attains an average anomaly detection F1-score of 93.6\%, representing an 8.8\% improvement over previous state-of-the-art methods. On Dataset A, our method improves upon the existing unsupervised state-of-the-art by 5.6 percentage points and surpasses the performance of existing multi-modal semi-supervised methods. Even without labelled data, our method achieves better results than semi-supervised methods.

On Dataset C, our model achieves improvements of 20.3 and 14.1 percentage points over the existing unsupervised and multi-modal semi-supervised state-of-the-art methods, respectively. Notably, Dataset C includes more metric items and parsed log templates, providing a more comprehensive depiction of the complex environment in real industrial scenarios. In such complex circumstances, asynchrony problems become more pronounced, making it difficult for traditional methods to sufficiently capture deep correlations between multi-modal data. This further validates the effectiveness of FFAD in handling complex data. FFAD effectively addresses the challenges posed by asynchrony by constructing a fully connected fusion graph that unifies the representation of each data point from different modalities.

Furthermore, our unsupervised method outperforms the semi-supervised method Hades on all three datasets. This may be because semi-supervised methods like Hades are susceptible to sparsely labelled data and have difficulty adapting to unseen anomalies. In contrast, our unsupervised method learns the implicit patterns of multi-modal data more comprehensively through reconstruction tasks unaffected by the scarcity of labelled data. This grants FFAD stronger generalization capabilities in detecting new anomalies, leading to better performance.

FFAD also demonstrates excellent recall performance. On Datasets B and C, our method achieves a recall rate of 100\%. This is because FFAD can effectively capture deep correlations between multi-modal data, comprehensively identifying potential anomalous signals and thus is less likely to miss anomalies. Although this may result in more false positives, we believe that in anomaly detection tasks, missing fewer anomalies is more important. In real industrial scenarios, undetected anomalies can lead to serious consequences, including system crashes, service interruptions, and even economic losses. Therefore, ensuring a high recall rate to capture all possible anomalies is crucial for maintaining system reliability and stability. Meanwhile, FFAD maintains high precision while sustaining high recall, ranking second only to the best method on Datasets B and achieving the highest precision on Datasets A and C.

\subsection{RQ2: How do different modules contribute to FFAD?}

In this section, we assess the effectiveness of FFAD's three key components: Temporal Feature Retention (TFR), Fourier Frequency Focus (FFF), and Dual Noise Injection (DNI). Table \ref{tab:ablation_results} summarizes the results of the ablation experiments. After individually removing each component from the model, FFAD's performance showed varying degrees of degradation, with the F1-scores across the three datasets decreasing by an average of 5.5\%.

Removing the TFR module led to an average F1-score decrease of 5.2\%. The TFR module helps retain temporal features and enhances the model's capture of temporal information. When we removed the TFR module, the model could not fully utilize temporal information, resulting in decreased anomaly detection performance.

Removing the DNI module caused an average F1-score decrease of 5.6\%. The DNI module balances the learning between the log and metric modalities. When we removed the DNI module, the model's learning on the modality with less data became insufficient, affecting overall performance. Notably, removing the DNI module had the greatest impact on Dataset B. There are two possible reasons for this. First, Datasets A and C added noise to the logs, while Dataset B added noise to the metrics. Since logs inherently introduce some noise during parsing, adding extra noise to the logs is less effective than adding noise to the relatively stable metric data. The second reason will be explained in the next section, RQ3.

Removing the FFF module resulted in an average F1-score decrease of 3.5\%. The FFF strategy effectively alleviates the overgeneralization problem by focusing on the frequency components of normal patterns in the Fourier frequency domain. When we remove the FFF module, the model might reconstruct abnormal data well, reducing its ability to distinguish anomalies.

\subsection{RQ3: Does data from each modality contribute to FFAD?}

\begin{figure*}[htbp]
    \centering
    % 第一张子图
    \begin{subfigure}[t]{0.32\textwidth}
        \centering
        \includegraphics[width=\textwidth]{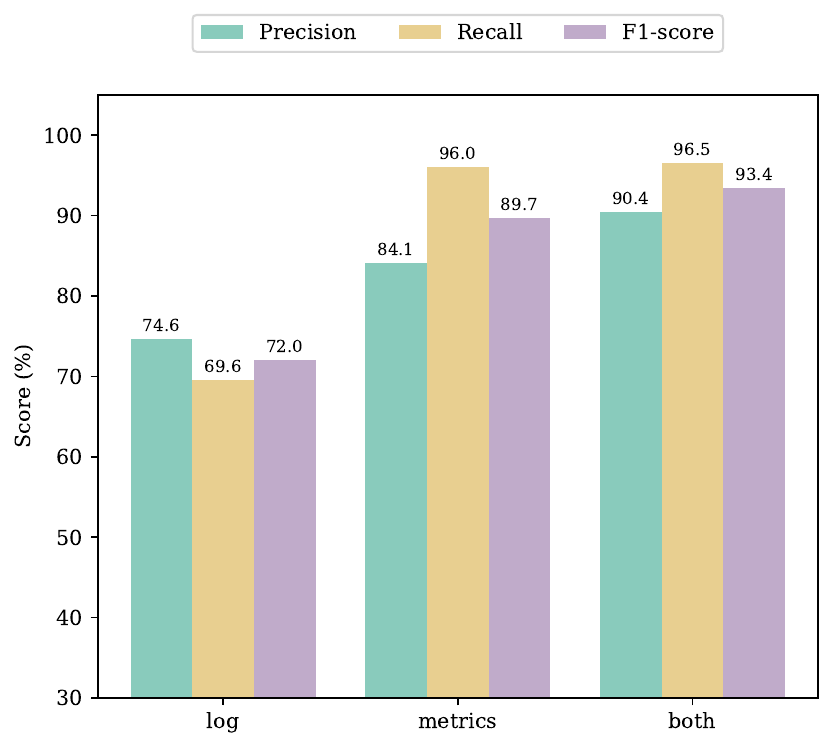}
        \caption{Dataset A}
        \label{fig: datasetA}
    \end{subfigure}
    \hfill
    % 第二张子图
    \begin{subfigure}[t]{0.32\textwidth}
        \centering
        \includegraphics[width=\textwidth]{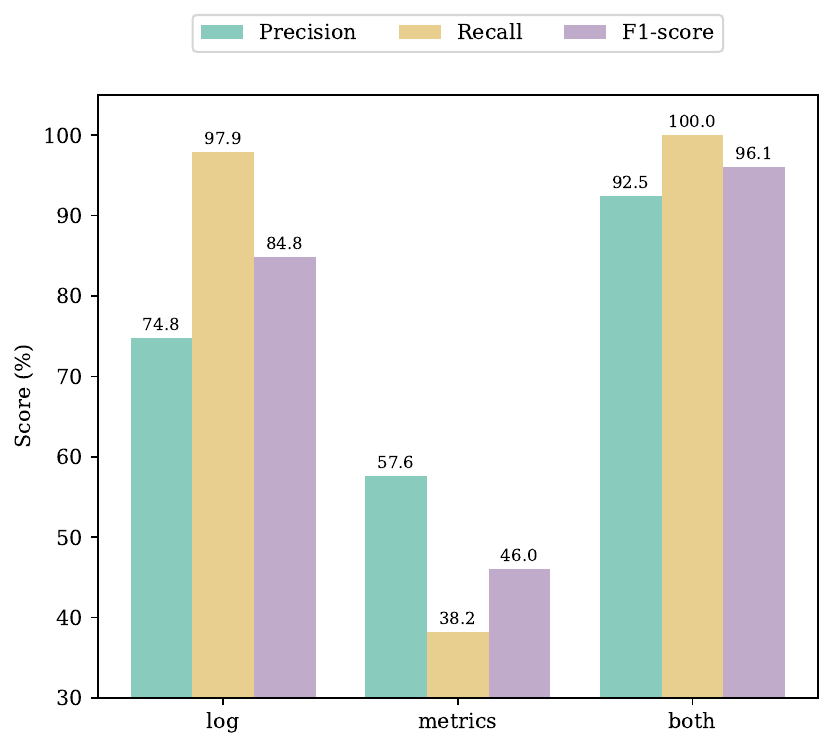}
        \caption{Dataset B}
        \label{fig: datasetB}
    \end{subfigure}
    \hfill
    % 第三张子图
    \begin{subfigure}[t]{0.32\textwidth}
        \centering
        \includegraphics[width=\textwidth]{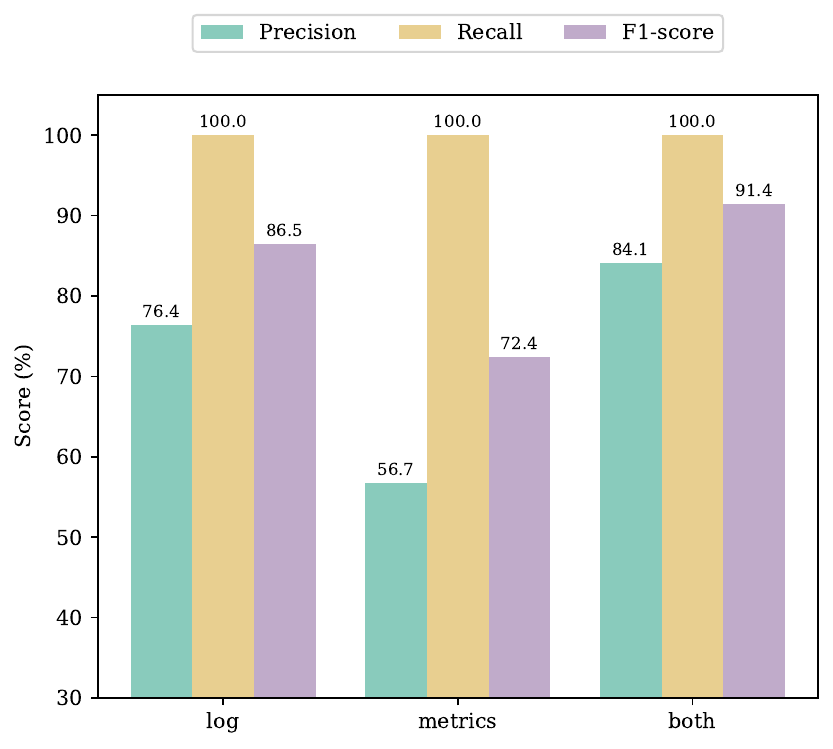}
        \caption{Dataset C}
        \label{fig: datasetC}
    \end{subfigure}
    \caption{Modality Data Ablation}
    \label{fig: mab}
    \vspace{-0.5cm}
\end{figure*} 

In this section, we evaluate the impact of different data modalities on the performance of FFAD. The results of our ablation studies for each data modality are shown in Figure \ref{fig: mab}. When different data modalities are ablated, FFAD's performance exhibits varying degrees of degradation across datasets. This is because different types of faults are typically reflected in different data modalities. The significant impact of ablating the DNI module on Dataset B, as observed in RQ2, can be explained here. The DNI module is crucial in balancing the learning between the log and metrics modalities. When the DNI module is removed, the model inadequately learns from the modality with less data, affecting overall performance. In Dataset B, the amount of metrics data is greater than the effective log data. Consequently, after removing the DNI module, the model tends to ignore the contribution of the log modality and over-relies on metrics data for judgments. However, as shown in Figure \ref{fig: datasetB}, when the log module is removed, FFAD's performance drops significantly, demonstrating that metrics data alone cannot detect most faults in Dataset B. Therefore, the DNI module has the greatest impact on Dataset B. In contrast, in Dataset A and Dataset C, both log and metrics data can independently detect many faults. Thus, removing the DNI module has a less significant impact on these two datasets than Dataset B. This further demonstrates that combining multiple data modalities enables a more comprehensive understanding of the system's state, leading to more accurate anomaly detection.

\subsection{RQ4: How is the effectiveness of the FFF mechanism visualized in FFAD}

\begin{figure}[h]
  \centering
  \begin{subfigure}[b]{0.225\textwidth}
    \centering
    \includegraphics[width=120px]{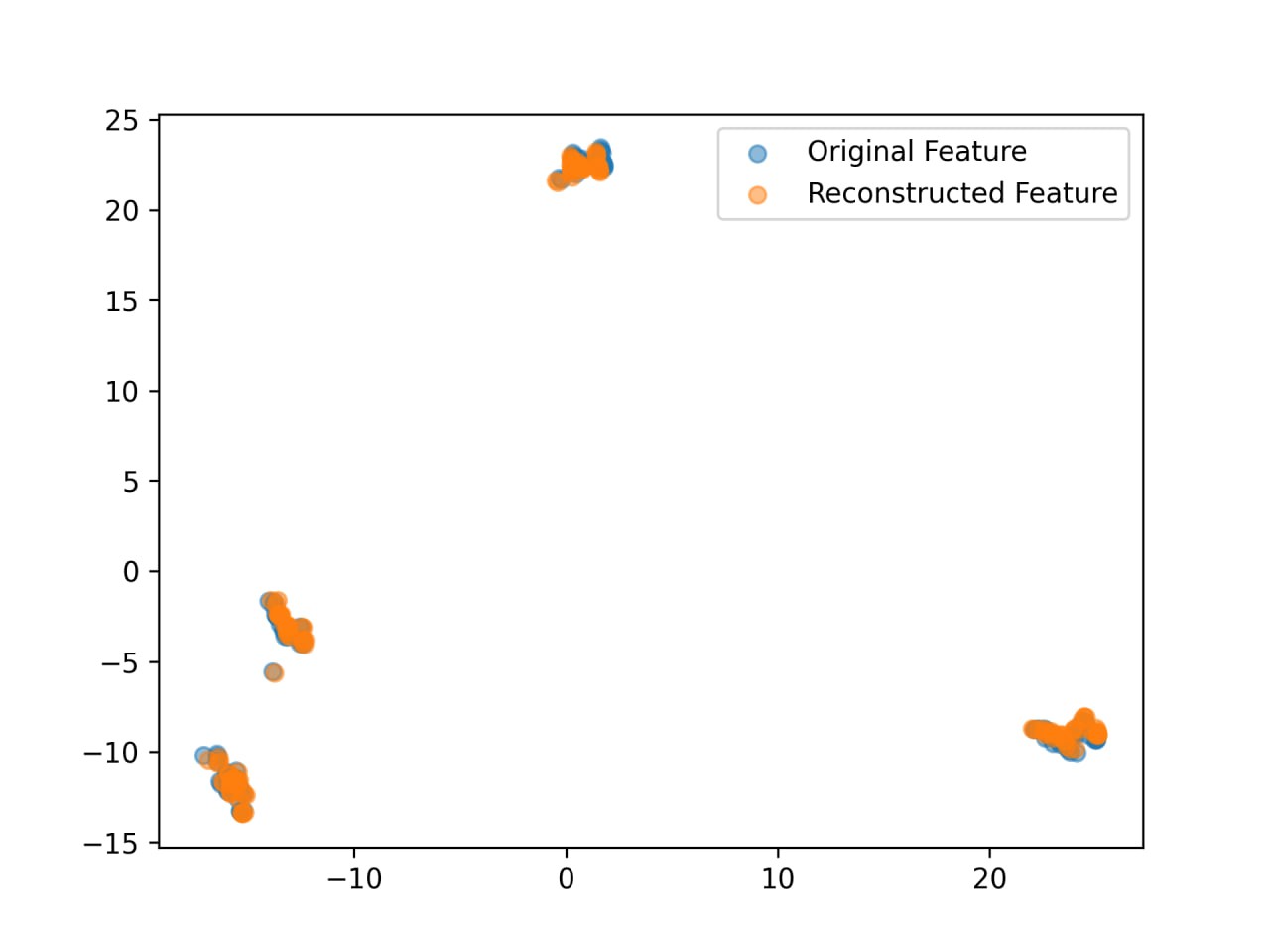}
    \caption{FFAD (w/o FFF)}
    \label{fig:origin}
    \Description{}
  \end{subfigure}
  \hfill
  \begin{subfigure}[b]{0.225\textwidth}
    \centering
    \includegraphics[width=120px]{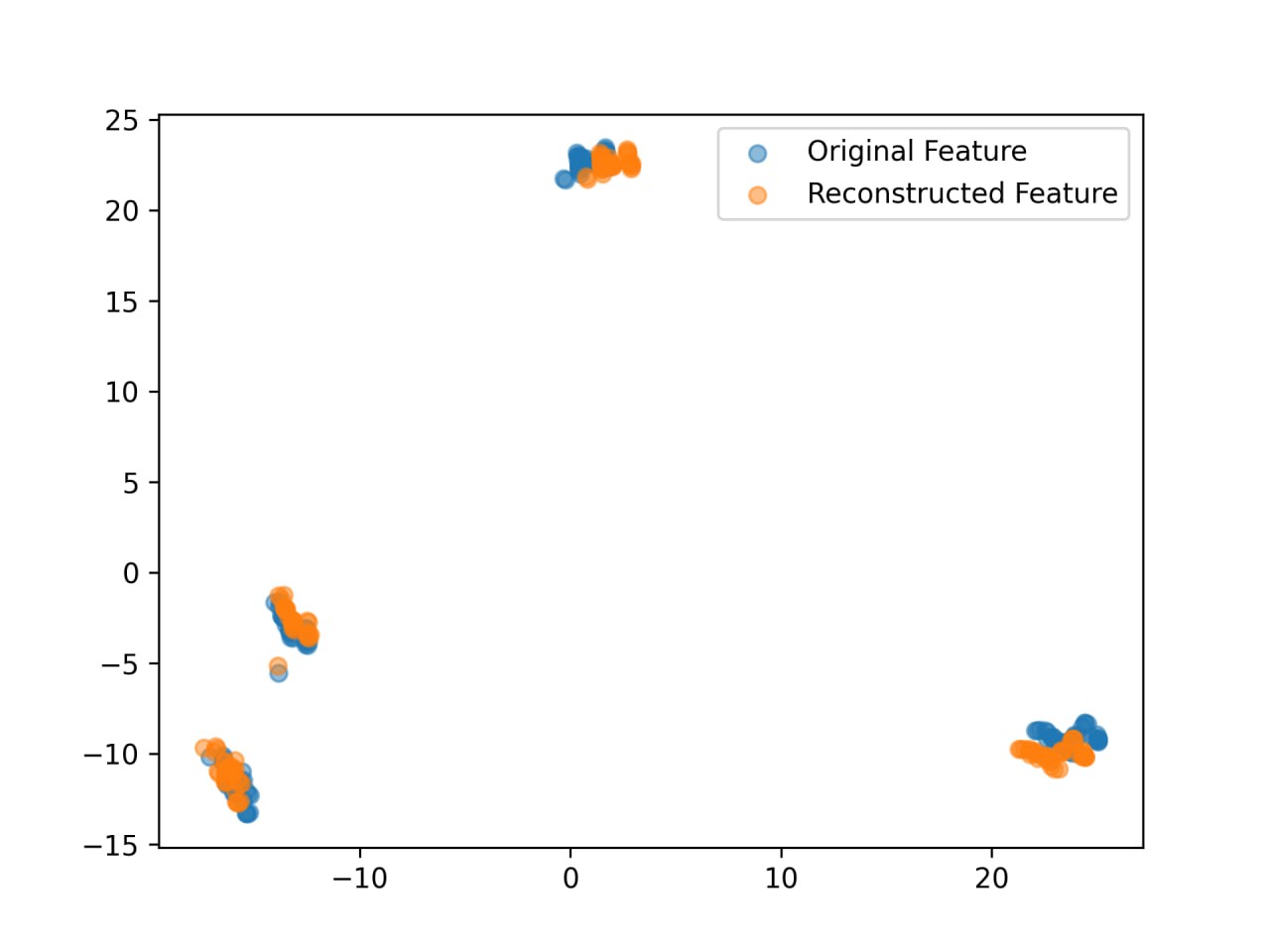}
    \caption{FFAD}
    \label{fig:recon}
    \Description{}
  \end{subfigure}
  
  \caption{Visualization Study}
  \vspace{-0.5cm}
\end{figure}

% \begin{figure}[ht]
%     \centering
%     \includegraphics[width=1.0\linewidth]{fff.jpg} 
%     % \caption{The precise association between logs and metrics within a time window.}
%     \label{fig:fff}
% \end{figure}

In this section, we explore the effectiveness of the Fourier Frequency Focus (FFF) mechanism in anomaly detection by visualizing the original and reconstructed features. To highlight the impact of the FFF mechanism, we selected a data segment containing many anomalies for visualization analysis. We conducted a comparative experiment: Figure \ref{fig:origin} presents the visualization results of the FFAD model converged without the FFF strategy. In contrast, Figure \ref{fig:recon} shows the visualization results of the FFAD model converged with the FFF strategy. Except for using the FFF mechanism, all other parameter settings were kept consistent to ensure a fair comparison. From Figure \ref{fig:origin}, we observe that without the FFF mechanism, the original features (blue points) and the reconstructed features (orange points) highly overlap. This indicates that the model not only reconstructs the normal features well but also overfits the anomalous features, making it difficult to distinguish anomalies effectively. In contrast, after adopting the FFF mechanism, as shown in Figure \ref{fig:recon}, most data points' original and reconstructed features still highly overlap, indicating that the reconstruction of normal patterns remains stable. However, the original and reconstructed features exhibit significant deviations in certain clusters. This demonstrates that the FFF mechanism successfully amplifies the reconstruction errors of anomalous features by dynamically adjusting frequency components with high signal intensity, while the reconstruction of normal patterns is not significantly affected. Therefore, the FFF mechanism allows the model to focus more on learning normal patterns. During the reconstruction process, this mechanism amplifies the reconstruction errors of anomalous points, thereby improving the ability to detect anomalies.

\section{CONCLUSION}

In this paper, we propose FFAD, a novel anomaly detection method for web services based on multi-modal data. To effectively detect anomalies in web service systems, it is essential to leverage both logs and metrics data. However, existing methods fail to address two key challenges: the precise association between logs and metrics, and the overgeneralization problem in reconstruction-based methods. FFAD addresses these challenges by employing a graph-based approach to achieve fine-grained association between logs and metrics, and introducing the Fourier Frequency Focus strategy to enhance the model's focus on normal patterns. Our method achieves an average F1-score improvement of 8.8\% over previous state-of-the-art methods on three datasets, including two real-world industrial datasets, demonstrating its effectiveness.

\section*{Acknowledgment}

This work was supported by National Key Laboratory of Data Space Technology and System.

%%
%% The next two lines define the bibliography style to be used, and
%% the bibliography file.
\bibliographystyle{ACM-Reference-Format}
\bibliography{sample-base}

%%% -*-BibTeX-*-
%%% Do NOT edit. File created by BibTeX with style
%%% ACM-Reference-Format-Journals [18-Jan-2012].

\begin{thebibliography}{31}

%%% ====================================================================
%%% NOTE TO THE USER: you can override these defaults by providing
%%% customized versions of any of these macros before the \bibliography
%%% command.  Each of them MUST provide its own final punctuation,
%%% except for \shownote{}, \showDOI{}, and \showURL{}.  The latter two
%%% do not use final punctuation, in order to avoid confusing it with
%%% the Web address.
%%%
%%% To suppress output of a particular field, define its macro to expand
%%% to an empty string, or better, \unskip, like this:
%%%
%%% \newcommand{\showDOI}[1]{\unskip}   % LaTeX syntax
%%%
%%% \def \showDOI #1{\unskip}           % plain TeX syntax
%%%
%%% ====================================================================

\ifx \showCODEN    \undefined \def \showCODEN     #1{\unskip}     \fi
\ifx \showDOI      \undefined \def \showDOI       #1{#1}\fi
\ifx \showISBNx    \undefined \def \showISBNx     #1{\unskip}     \fi
\ifx \showISBNxiii \undefined \def \showISBNxiii  #1{\unskip}     \fi
\ifx \showISSN     \undefined \def \showISSN      #1{\unskip}     \fi
\ifx \showLCCN     \undefined \def \showLCCN      #1{\unskip}     \fi
\ifx \shownote     \undefined \def \shownote      #1{#1}          \fi
\ifx \showarticletitle \undefined \def \showarticletitle #1{#1}   \fi
\ifx \showURL      \undefined \def \showURL       {\relax}        \fi
% The following commands are used for tagged output and should be
% invisible to TeX
\providecommand\bibfield[2]{#2}
\providecommand\bibinfo[2]{#2}
\providecommand\natexlab[1]{#1}
\providecommand\showeprint[2][]{arXiv:#2}

\bibitem[Breunig et~al\mbox{.}(2000)]%
        {breunig2000lof}
\bibfield{author}{\bibinfo{person}{Markus~M Breunig}, \bibinfo{person}{Hans-Peter Kriegel}, \bibinfo{person}{Raymond~T Ng}, {and} \bibinfo{person}{J{\"o}rg Sander}.} \bibinfo{year}{2000}\natexlab{}.
\newblock \showarticletitle{LOF: identifying density-based local outliers}. In \bibinfo{booktitle}{\emph{Proceedings of the 2000 ACM SIGMOD international conference on Management of data}}. \bibinfo{pages}{93--104}.
\newblock


\bibitem[Chen et~al\mbox{.}(2024a)]%
        {chen2024lara}
\bibfield{author}{\bibinfo{person}{Feiyi Chen}, \bibinfo{person}{Zhen Qin}, \bibinfo{person}{Mengchu Zhou}, \bibinfo{person}{Yingying Zhang}, \bibinfo{person}{Shuiguang Deng}, \bibinfo{person}{Lunting Fan}, \bibinfo{person}{Guansong Pang}, {and} \bibinfo{person}{Qingsong Wen}.} \bibinfo{year}{2024}\natexlab{a}.
\newblock \showarticletitle{LARA: A Light and Anti-overfitting Retraining Approach for Unsupervised Time Series Anomaly Detection}. In \bibinfo{booktitle}{\emph{Proceedings of the ACM on Web Conference 2024}}. \bibinfo{pages}{4138--4149}.
\newblock


\bibitem[Chen et~al\mbox{.}(2024b)]%
        {chen2024learning}
\bibfield{author}{\bibinfo{person}{Feiyi Chen}, \bibinfo{person}{Yingying Zhang}, \bibinfo{person}{Zhen Qin}, \bibinfo{person}{Lunting Fan}, \bibinfo{person}{Renhe Jiang}, \bibinfo{person}{Yuxuan Liang}, \bibinfo{person}{Qingsong Wen}, {and} \bibinfo{person}{Shuiguang Deng}.} \bibinfo{year}{2024}\natexlab{b}.
\newblock \showarticletitle{Learning Multi-Pattern Normalities in the Frequency Domain for Efficient Time Series Anomaly Detection}. In \bibinfo{booktitle}{\emph{2024 IEEE 40th International Conference on Data Engineering (ICDE)}}. IEEE, \bibinfo{pages}{747--760}.
\newblock


\bibitem[Du et~al\mbox{.}(2017)]%
        {du2017deeplog}
\bibfield{author}{\bibinfo{person}{Min Du}, \bibinfo{person}{Feifei Li}, \bibinfo{person}{Guineng Zheng}, {and} \bibinfo{person}{Vivek Srikumar}.} \bibinfo{year}{2017}\natexlab{}.
\newblock \showarticletitle{Deeplog: Anomaly detection and diagnosis from system logs through deep learning}. In \bibinfo{booktitle}{\emph{Proceedings of the 2017 ACM SIGSAC conference on computer and communications security}}. \bibinfo{pages}{1285--1298}.
\newblock


\bibitem[Fu et~al\mbox{.}(2023)]%
        {fu2023mlog}
\bibfield{author}{\bibinfo{person}{Yuanyuan Fu}, \bibinfo{person}{Kun Liang}, {and} \bibinfo{person}{Jian Xu}.} \bibinfo{year}{2023}\natexlab{}.
\newblock \showarticletitle{MLog: Mogrifier LSTM-based log anomaly detection approach using semantic representation}.
\newblock \bibinfo{journal}{\emph{IEEE Transactions on Services Computing}} \bibinfo{volume}{16}, \bibinfo{number}{5} (\bibinfo{year}{2023}), \bibinfo{pages}{3537--3549}.
\newblock


\bibitem[Guan et~al\mbox{.}(2023)]%
        {guan2023grasped}
\bibfield{author}{\bibinfo{person}{Wei Guan}, \bibinfo{person}{Jian Cao}, \bibinfo{person}{Yang Gu}, {and} \bibinfo{person}{Shiyou Qian}.} \bibinfo{year}{2023}\natexlab{}.
\newblock \showarticletitle{Grasped: A gru-ae network based multi-perspective business process anomaly detection model}.
\newblock \bibinfo{journal}{\emph{IEEE Transactions on Services Computing}} \bibinfo{volume}{16}, \bibinfo{number}{5} (\bibinfo{year}{2023}), \bibinfo{pages}{3412--3424}.
\newblock


\bibitem[He et~al\mbox{.}(2017)]%
        {he2017drain}
\bibfield{author}{\bibinfo{person}{Pinjia He}, \bibinfo{person}{Jieming Zhu}, \bibinfo{person}{Zibin Zheng}, {and} \bibinfo{person}{Michael~R Lyu}.} \bibinfo{year}{2017}\natexlab{}.
\newblock \showarticletitle{Drain: An online log parsing approach with fixed depth tree}. In \bibinfo{booktitle}{\emph{2017 IEEE international conference on web services (ICWS)}}. IEEE, \bibinfo{pages}{33--40}.
\newblock


\bibitem[Jie et~al\mbox{.}(2024)]%
        {jie2024disentangled}
\bibfield{author}{\bibinfo{person}{Xin Jie}, \bibinfo{person}{Xixi Zhou}, \bibinfo{person}{Chanfei Su}, \bibinfo{person}{Zijun Zhou}, \bibinfo{person}{Yuqing Yuan}, \bibinfo{person}{Jiajun Bu}, {and} \bibinfo{person}{Haishuai Wang}.} \bibinfo{year}{2024}\natexlab{}.
\newblock \showarticletitle{Disentangled Anomaly Detection For Multivariate Time Series}. In \bibinfo{booktitle}{\emph{Companion Proceedings of the ACM on Web Conference 2024}}. \bibinfo{pages}{931--934}.
\newblock


\bibitem[Joulin(2016)]%
        {joulin2016fasttext}
\bibfield{author}{\bibinfo{person}{Armand Joulin}.} \bibinfo{year}{2016}\natexlab{}.
\newblock \showarticletitle{Fasttext. zip: Compressing text classification models}.
\newblock \bibinfo{journal}{\emph{arXiv preprint arXiv:1612.03651}} (\bibinfo{year}{2016}).
\newblock


\bibitem[Lee et~al\mbox{.}(2023)]%
        {lee2023heterogeneous}
\bibfield{author}{\bibinfo{person}{Cheryl Lee}, \bibinfo{person}{Tianyi Yang}, \bibinfo{person}{Zhuangbin Chen}, \bibinfo{person}{Yuxin Su}, \bibinfo{person}{Yongqiang Yang}, {and} \bibinfo{person}{Michael~R Lyu}.} \bibinfo{year}{2023}\natexlab{}.
\newblock \showarticletitle{Heterogeneous anomaly detection for software systems via semi-supervised cross-modal attention}. In \bibinfo{booktitle}{\emph{2023 IEEE/ACM 45th International Conference on Software Engineering (ICSE)}}. IEEE, \bibinfo{pages}{1724--1736}.
\newblock


\bibitem[Lin et~al\mbox{.}(2016)]%
        {lin2016log}
\bibfield{author}{\bibinfo{person}{Qingwei Lin}, \bibinfo{person}{Hongyu Zhang}, \bibinfo{person}{Jian-Guang Lou}, \bibinfo{person}{Yu Zhang}, {and} \bibinfo{person}{Xuewei Chen}.} \bibinfo{year}{2016}\natexlab{}.
\newblock \showarticletitle{Log clustering based problem identification for online service systems}. In \bibinfo{booktitle}{\emph{Proceedings of the 38th International Conference on Software Engineering Companion}}. \bibinfo{pages}{102--111}.
\newblock


\bibitem[Liu et~al\mbox{.}(2024)]%
        {liu2024uac}
\bibfield{author}{\bibinfo{person}{Hongyi Liu}, \bibinfo{person}{Xiaosong Huang}, \bibinfo{person}{Mengxi Jia}, \bibinfo{person}{Tong Jia}, \bibinfo{person}{Jing Han}, \bibinfo{person}{Ying Li}, {and} \bibinfo{person}{Zhonghai Wu}.} \bibinfo{year}{2024}\natexlab{}.
\newblock \showarticletitle{UAC-AD: Unsupervised Adversarial Contrastive Learning for Anomaly Detection on Multi-modal Data in Microservice Systems}.
\newblock \bibinfo{journal}{\emph{IEEE Transactions on Services Computing}} (\bibinfo{year}{2024}).
\newblock


\bibitem[{Locust Authors}({[n.\,d.]})]%
        {locust}
\bibfield{author}{\bibinfo{person}{{Locust Authors}}.} \bibinfo{year}{[n.\,d.]}\natexlab{}.
\newblock \bibinfo{title}{Locust}.
\newblock \bibinfo{howpublished}{\url{https://locust.io}}.
\newblock


\bibitem[Ma et~al\mbox{.}(2021)]%
        {ma2021jump}
\bibfield{author}{\bibinfo{person}{Minghua Ma}, \bibinfo{person}{Shenglin Zhang}, \bibinfo{person}{Junjie Chen}, \bibinfo{person}{Jim Xu}, \bibinfo{person}{Haozhe Li}, \bibinfo{person}{Yongliang Lin}, \bibinfo{person}{Xiaohui Nie}, \bibinfo{person}{Bo Zhou}, \bibinfo{person}{Yong Wang}, {and} \bibinfo{person}{Dan Pei}.} \bibinfo{year}{2021}\natexlab{}.
\newblock \showarticletitle{$\{$Jump-Starting$\}$ multivariate time series anomaly detection for online service systems}. In \bibinfo{booktitle}{\emph{2021 USENIX Annual Technical Conference (USENIX ATC 21)}}. \bibinfo{pages}{413--426}.
\newblock


\bibitem[Park et~al\mbox{.}(2018)]%
        {park2018multimodal}
\bibfield{author}{\bibinfo{person}{Daehyung Park}, \bibinfo{person}{Yuuna Hoshi}, {and} \bibinfo{person}{Charles~C Kemp}.} \bibinfo{year}{2018}\natexlab{}.
\newblock \showarticletitle{A multimodal anomaly detector for robot-assisted feeding using an lstm-based variational autoencoder}.
\newblock \bibinfo{journal}{\emph{IEEE Robotics and Automation Letters}} \bibinfo{volume}{3}, \bibinfo{number}{3} (\bibinfo{year}{2018}), \bibinfo{pages}{1544--1551}.
\newblock


\bibitem[Park et~al\mbox{.}(2020)]%
        {park2020learning}
\bibfield{author}{\bibinfo{person}{Hyunjong Park}, \bibinfo{person}{Jongyoun Noh}, {and} \bibinfo{person}{Bumsub Ham}.} \bibinfo{year}{2020}\natexlab{}.
\newblock \showarticletitle{Learning memory-guided normality for anomaly detection}. In \bibinfo{booktitle}{\emph{Proceedings of the IEEE/CVF conference on computer vision and pattern recognition}}. \bibinfo{pages}{14372--14381}.
\newblock


\bibitem[Shen et~al\mbox{.}(2020)]%
        {shen2020timeseries}
\bibfield{author}{\bibinfo{person}{Lifeng Shen}, \bibinfo{person}{Zhuocong Li}, {and} \bibinfo{person}{James Kwok}.} \bibinfo{year}{2020}\natexlab{}.
\newblock \showarticletitle{Timeseries anomaly detection using temporal hierarchical one-class network}.
\newblock \bibinfo{journal}{\emph{Advances in Neural Information Processing Systems}}  \bibinfo{volume}{33} (\bibinfo{year}{2020}), \bibinfo{pages}{13016--13026}.
\newblock


\bibitem[Song et~al\mbox{.}(2023)]%
        {song2023memto}
\bibfield{author}{\bibinfo{person}{Junho Song}, \bibinfo{person}{Keonwoo Kim}, \bibinfo{person}{Jeonglyul Oh}, {and} \bibinfo{person}{Sungzoon Cho}.} \bibinfo{year}{2023}\natexlab{}.
\newblock \showarticletitle{Memto: Memory-guided transformer for multivariate time series anomaly detection}.
\newblock \bibinfo{journal}{\emph{Advances in Neural Information Processing Systems}}  \bibinfo{volume}{36} (\bibinfo{year}{2023}), \bibinfo{pages}{57947--57963}.
\newblock


\bibitem[Su et~al\mbox{.}(2019)]%
        {su2019robust}
\bibfield{author}{\bibinfo{person}{Ya Su}, \bibinfo{person}{Youjian Zhao}, \bibinfo{person}{Chenhao Niu}, \bibinfo{person}{Rong Liu}, \bibinfo{person}{Wei Sun}, {and} \bibinfo{person}{Dan Pei}.} \bibinfo{year}{2019}\natexlab{}.
\newblock \showarticletitle{Robust anomaly detection for multivariate time series through stochastic recurrent neural network}. In \bibinfo{booktitle}{\emph{Proceedings of the 25th ACM SIGKDD international conference on knowledge discovery \& data mining}}. \bibinfo{pages}{2828--2837}.
\newblock


\bibitem[Tan et~al\mbox{.}(2024)]%
        {tan2024air}
\bibfield{author}{\bibinfo{person}{Yuanzheng Tan}, \bibinfo{person}{Qing Li}, \bibinfo{person}{Junkun Peng}, \bibinfo{person}{Zhenhui Yuan}, {and} \bibinfo{person}{Yong Jiang}.} \bibinfo{year}{2024}\natexlab{}.
\newblock \showarticletitle{Air-CAD: Edge-Assisted Multi-Drone Network for Real-time Crowd Anomaly Detection}. In \bibinfo{booktitle}{\emph{Proceedings of the ACM on Web Conference 2024}}. \bibinfo{pages}{2817--2825}.
\newblock


\bibitem[Tang et~al\mbox{.}(2002)]%
        {tang2002enhancing}
\bibfield{author}{\bibinfo{person}{Jian Tang}, \bibinfo{person}{Zhixiang Chen}, \bibinfo{person}{Ada Wai-Chee Fu}, {and} \bibinfo{person}{David~W Cheung}.} \bibinfo{year}{2002}\natexlab{}.
\newblock \showarticletitle{Enhancing effectiveness of outlier detections for low density patterns}. In \bibinfo{booktitle}{\emph{Advances in Knowledge Discovery and Data Mining: 6th Pacific-Asia Conference, PAKDD 2002 Taipei, Taiwan, May 6--8, 2002 Proceedings 6}}. Springer, \bibinfo{pages}{535--548}.
\newblock


\bibitem[Tax and Duin(2004)]%
        {tax2004support}
\bibfield{author}{\bibinfo{person}{David~MJ Tax} {and} \bibinfo{person}{Robert~PW Duin}.} \bibinfo{year}{2004}\natexlab{}.
\newblock \showarticletitle{Support vector data description}.
\newblock \bibinfo{journal}{\emph{Machine learning}}  \bibinfo{volume}{54} (\bibinfo{year}{2004}), \bibinfo{pages}{45--66}.
\newblock


\bibitem[Xu(2010)]%
        {xu2010system}
\bibfield{author}{\bibinfo{person}{Wei Xu}.} \bibinfo{year}{2010}\natexlab{}.
\newblock \bibinfo{booktitle}{\emph{System problem detection by mining console logs}}.
\newblock \bibinfo{publisher}{University of California, Berkeley}.
\newblock


\bibitem[Yi et~al\mbox{.}(2024)]%
        {yi2024fouriergnn}
\bibfield{author}{\bibinfo{person}{Kun Yi}, \bibinfo{person}{Qi Zhang}, \bibinfo{person}{Wei Fan}, \bibinfo{person}{Hui He}, \bibinfo{person}{Liang Hu}, \bibinfo{person}{Pengyang Wang}, \bibinfo{person}{Ning An}, \bibinfo{person}{Longbing Cao}, {and} \bibinfo{person}{Zhendong Niu}.} \bibinfo{year}{2024}\natexlab{}.
\newblock \showarticletitle{FourierGNN: Rethinking multivariate time series forecasting from a pure graph perspective}.
\newblock \bibinfo{journal}{\emph{Advances in Neural Information Processing Systems}}  \bibinfo{volume}{36} (\bibinfo{year}{2024}).
\newblock


\bibitem[Zhan et~al\mbox{.}(2022)]%
        {zhan2022stgat}
\bibfield{author}{\bibinfo{person}{Jun Zhan}, \bibinfo{person}{Siqi Wang}, \bibinfo{person}{Xiandong Ma}, \bibinfo{person}{Chengkun Wu}, \bibinfo{person}{Canqun Yang}, \bibinfo{person}{Detian Zeng}, {and} \bibinfo{person}{Shilin Wang}.} \bibinfo{year}{2022}\natexlab{}.
\newblock \showarticletitle{Stgat-mad: Spatial-temporal graph attention network for multivariate time series anomaly detection}. In \bibinfo{booktitle}{\emph{ICASSP 2022-2022 IEEE International Conference on Acoustics, Speech and Signal Processing (ICASSP)}}. IEEE, \bibinfo{pages}{3568--3572}.
\newblock


\bibitem[Zhang et~al\mbox{.}(2019)]%
        {zhang2019robust}
\bibfield{author}{\bibinfo{person}{Xu Zhang}, \bibinfo{person}{Yong Xu}, \bibinfo{person}{Qingwei Lin}, \bibinfo{person}{Bo Qiao}, \bibinfo{person}{Hongyu Zhang}, \bibinfo{person}{Yingnong Dang}, \bibinfo{person}{Chunyu Xie}, \bibinfo{person}{Xinsheng Yang}, \bibinfo{person}{Qian Cheng}, \bibinfo{person}{Ze Li}, {et~al\mbox{.}}} \bibinfo{year}{2019}\natexlab{}.
\newblock \showarticletitle{Robust log-based anomaly detection on unstable log data}. In \bibinfo{booktitle}{\emph{Proceedings of the 2019 27th ACM joint meeting on European software engineering conference and symposium on the foundations of software engineering}}. \bibinfo{pages}{807--817}.
\newblock


\bibitem[Zhao et~al\mbox{.}(2020)]%
        {zhao2020multivariate}
\bibfield{author}{\bibinfo{person}{Hang Zhao}, \bibinfo{person}{Yujing Wang}, \bibinfo{person}{Juanyong Duan}, \bibinfo{person}{Congrui Huang}, \bibinfo{person}{Defu Cao}, \bibinfo{person}{Yunhai Tong}, \bibinfo{person}{Bixiong Xu}, \bibinfo{person}{Jing Bai}, \bibinfo{person}{Jie Tong}, {and} \bibinfo{person}{Qi Zhang}.} \bibinfo{year}{2020}\natexlab{}.
\newblock \showarticletitle{Multivariate time-series anomaly detection via graph attention network}. In \bibinfo{booktitle}{\emph{2020 IEEE international conference on data mining (ICDM)}}. IEEE, \bibinfo{pages}{841--850}.
\newblock


\bibitem[Zhao et~al\mbox{.}(2024)]%
        {zhao2024weakly}
\bibfield{author}{\bibinfo{person}{Haihong Zhao}, \bibinfo{person}{Chenyi Zi}, \bibinfo{person}{Yang Liu}, \bibinfo{person}{Chen Zhang}, \bibinfo{person}{Yan Zhou}, {and} \bibinfo{person}{Jia Li}.} \bibinfo{year}{2024}\natexlab{}.
\newblock \showarticletitle{Weakly Supervised Anomaly Detection via Knowledge-Data Alignment}. In \bibinfo{booktitle}{\emph{Proceedings of the ACM on Web Conference 2024}}. \bibinfo{pages}{4083--4094}.
\newblock


\bibitem[Zhao et~al\mbox{.}(2021)]%
        {zhao2021identifying}
\bibfield{author}{\bibinfo{person}{Nengwen Zhao}, \bibinfo{person}{Junjie Chen}, \bibinfo{person}{Zhaoyang Yu}, \bibinfo{person}{Honglin Wang}, \bibinfo{person}{Jiesong Li}, \bibinfo{person}{Bin Qiu}, \bibinfo{person}{Hongyu Xu}, \bibinfo{person}{Wenchi Zhang}, \bibinfo{person}{Kaixin Sui}, {and} \bibinfo{person}{Dan Pei}.} \bibinfo{year}{2021}\natexlab{}.
\newblock \showarticletitle{Identifying bad software changes via multimodal anomaly detection for online service systems}. In \bibinfo{booktitle}{\emph{Proceedings of the 29th ACM Joint Meeting on European Software Engineering Conference and Symposium on the Foundations of Software Engineering}}. \bibinfo{pages}{527--539}.
\newblock


\bibitem[Zhou et~al\mbox{.}(2018)]%
        {train-ticket}
\bibfield{author}{\bibinfo{person}{Xiang Zhou}, \bibinfo{person}{Xin Peng}, \bibinfo{person}{Tao Xie}, \bibinfo{person}{Jun Sun}, \bibinfo{person}{Chenjie Xu}, \bibinfo{person}{Chao Ji}, {and} \bibinfo{person}{Wenyun Zhao}.} \bibinfo{year}{2018}\natexlab{}.
\newblock \showarticletitle{Benchmarking microservice systems for software engineering research}. In \bibinfo{booktitle}{\emph{Proceedings of the 40th International Conference on Software Engineering: Companion Proceeedings}}. \bibinfo{pages}{323--324}.
\newblock


\bibitem[Zong et~al\mbox{.}(2018)]%
        {zong2018deep}
\bibfield{author}{\bibinfo{person}{Bo Zong}, \bibinfo{person}{Qi Song}, \bibinfo{person}{Martin~Renqiang Min}, \bibinfo{person}{Wei Cheng}, \bibinfo{person}{Cristian Lumezanu}, \bibinfo{person}{Daeki Cho}, {and} \bibinfo{person}{Haifeng Chen}.} \bibinfo{year}{2018}\natexlab{}.
\newblock \showarticletitle{Deep autoencoding gaussian mixture model for unsupervised anomaly detection}. In \bibinfo{booktitle}{\emph{International conference on learning representations}}.
\newblock


\end{thebibliography}

\end{document}